\documentclass[prd,reprint,nofootinbib,showpacs]{revtex4-1}
\usepackage{amsmath, amssymb,bm,xcolor,graphicx}
\usepackage[colorlinks=true,citecolor=blue]{hyperref}
\usepackage[T1]{fontenc}

\DeclareMathOperator{\sinc}{sinc}

\begin{document}
\title{Generating exact solutions to Einstein's equation using linearized approximations}
\author{Abraham I. Harte}
\email{harte@aei.mpg.de}
\author{Justin Vines}
\email{justin.vines@aei.mpg.de}
\affiliation{
 Max-Planck-Institut f\"ur Gravitationsphysik, Albert-Einstein-Institut
 \\ Am M\"uhlenberg 1, 14476 Golm, Germany.}

\begin{abstract}
We show that certain solutions to the linearized Einstein equation can---by the application of a particular type of linearized gauge transformation---be straightforwardly transformed into solutions of the exact Einstein equation. In cases with nontrivial matter content, the exact stress-energy tensor of the transformed metric has the same eigenvalues and eigenvectors as the linearized stress-energy tensor of the initial approximation. When our gauge exists, the tensorial structure of transformed metric perturbations identically eliminates all nonlinearities in Einstein's equation. As examples, we derive the exact Kerr and gravitational plane wave metrics from standard harmonic-gauge approximations. 
\end{abstract}

\maketitle

\vskip 1pc

\section{Introduction}

Perturbation theory in general relativity has by now reached an impressive degree of development. The context in which it is perhaps most mature is the post-Newtonian approximation, which assumes small perturbations to a Minkowski metric, weak stresses, and slow motion---at least in regions exterior to any compact bodies which may be present \cite{FutamaseLRR, BlanchetLRR, PoissonWill}. With these assumptions, one essentially\footnote{Non-systematic aspects remain if one insists on using point particles in the post-Newtonian approximation. These objects do not make sense in any generic way in standard general relativity \cite{GerochTraschen}, and forcing them into the approximation is accomplished in practice \cite{BlanchetLRR} by supplementing the theory with additional structure, namely regularization procedures. While there are often physical motivations for these procedures, ambiguities remain and a systematic understanding from first principles is lacking.} has a systematic algorithm which can be used to generate as many terms as desired in approximate solutions to Einstein's equation. While considerable calculational effort is required to obtain each new term, many of the underlying conceptual issues have now been resolved. Despite this, there are various senses in which the resulting approximations might not be optimal: They may fail to incorporate all available information, the chosen variables might be poorly adapted to the physical degrees of freedom, and so on. Similar comments apply, of course, to perturbative methods used in almost every branch of physics.

For these reasons and others, it can be useful to apply ``convergence acceleration'' or ``resummation'' techniques, which attempt to improve the overall quality of an approximation by combining information from one or more systematic expansions together with physically-motivated guesses for the behavior of the higher-order terms. Some such techniques---Richardson extrapolation, Shanks transforms, Borel summation, Pad\'{e} approximants, and so on---adopt a generic approach, taking into account mainly the mathematical structure of whichever partial sums happen to be available \cite{BenderOrszag, ResummationRev}. Whether or not any of these techniques might be useful in a particular problem can depend on the presence and locations of poles, whether the higher-order coefficients are monotonic or oscillatory, or other qualitative features. They do not, however, depend on details of the underlying physics. As a consequence, these types of techniques can improve approximations in a wide range of physically-dissimilar problems.

Other resummation techniques are more specialized, attempting to directly take advantage of the detailed physical or mathematical features specific to the problem at hand. Perhaps the most developed of these techniques in general relativity are incorporated into the effective-one-body approach \cite{EOB,DamourReview}, which uses some of the aforementioned generic resummation techniques, physical arguments, and information from the extreme-mass-ratio and post-Newtonian approximations to motivate a scaffold upon which a ``complete'' solution to the black hole two-body problem might be built. One finds, e.g., that at least at low orders, the post-Newtonian Hamiltonians which describe the conservative dynamics of two black holes simplify considerably after applying the resummations and variable redefinitions associated with the effective-one-body approach. 

These results and others provide strong evidence that existing systematic approximation schemes in general relativity can be improved. While various techniques have already been applied in different contexts, it is mostly unclear why or when any of these techniques work. A systematic understanding is severely lacking, and progress is made largely by \textit{ad hoc} experimentation. It appears useful at this stage to consolidate what has been learned, and to attempt a more systematic approach in which improved approximation techniques are obtained directly from the underlying theory. The goal of this paper is to present one new result which might eventually be incorporated into the development of just such a formalism.

Our discussion can be interpreted in the context of gauge choice.  Existing approaches to perturbation theory in general relativity adopt particular gauges mainly to simplify calculations or to provide more transparent physical interpretations, but not to improve accuracy; gauges are typically viewed as having no true physical content. While this is indeed true for exact solutions, it is not necessarily so in perturbation theory. Two approximations which are gauge-equivalent at a given perturbative order might have higher-order errors which are quite different. It would clearly be advantageous to identify a gauge which systematically minimizes these higher-order errors.

This paper provides just such a gauge, at least for a certain class of metrics. More precisely, we show that when it exists, the application of a particular first-order gauge transformation eliminates \textit{all} higher-order errors; first-order approximations are transformed into exact solutions to Einstein's equation. While the technique applies only to very special metrics, it does so for some of the most important examples which are known. In particular, it works for Kerr black holes. It does not apply directly for the more complicated example of a black hole binary, although much of the strong-field region in such a system is ``close'' to Kerr, and one might expect that the majority of the strong-field behavior---which is otherwise difficult to capture in perturbation theory---might be taken into account via a scheme inspired by the one described here.

Interpreted somewhat differently, our result draws attention to the fact that small parameters are not the only structures which can simplify approximations in general relativity; the details of a perturbation's tensorial structure can also be essential. The Lorentzian nature of relativistic metrics allows, for example, the presence of nontrivial perturbations which square to zero. These behave much more simply in Einstein's equation than more general perturbations, and are precisely the cases highlighted here. It appears fortuitous that  perturbations with this mathematical structure arise not merely as curiosities, but also as descriptions for some of the most physically-important spacetimes which are known.

Sect. \ref{Sect:PertTheory} reviews linearized perturbation theory in general relativity, while Sect. \ref{Sect:KS} recalls several results essential to our discussion regarding the so-called Kerr-Schild class of metrics \cite{KerrSchild, ExactSolns}. Sect. \ref{Sect:genSoln} then describes how to generate exact solutions by applying gauge transformations to approximate solutions, how the matter content in the approximate and transformed metrics is related, and how conservation laws work in this context. Sect. \ref{Sect:Examples} provides some simple examples of this technique, generating the exact Kerr and gravitational plane wave solutions, as well as static, spherically-symmetric metrics. Lastly, Sect. \ref{Sect:Discussion} speculates on how our results might be incorporated into a more general perturbative formalism which is no longer restricted to special types of perturbations.

Our sign and index conventions follow those of Wald \cite{Wald}. We use, for example, $a,b,\ldots$ for abstract indices, $\mu,\nu,\ldots$ for spacetimes coordinate indices, and $i,j,\ldots$ for spatial coordinate indices. Riemann tensors are defined such that $2 \nabla_{[a} \nabla_{b]} \omega_c = R_{abc}{}^{d} \omega_d$ for any 1-form $\omega_a$. We often work with two or more metrics, but raise and lower indices only using background metrics. Also note that although many of our results are easily generalized, we restrict for concreteness only to four-dimensional spacetimes.

\section{Linearized general relativity}
\label{Sect:PertTheory}

Our first step is to review perturbation theory in general relativity in its simplest, linearized form. Consider some well-behaved background metric $\bar{g}_{ab}$, and the 1-parameter family of deformations
\begin{equation}
  g^{(\epsilon)}_{ab} = \bar{g}_{ab} + \epsilon h_{ab},
  \label{gFamily}
\end{equation}
where $h_{ab}$ is fixed and $\epsilon \geq 0$ is an arbitrary parameter which is assumed to be sufficiently close to zero that $g_{ab}^{(\epsilon)}$ remains invertible in all regions of interest. We now allow for a (possibly vanishing) cosmological constant $\Lambda$ and use Einstein's equation
\begin{equation}
  R^{(\epsilon)}_{ab} - \frac{1}{2} g^{(\epsilon)}_{ab} R^{(\epsilon)} + \Lambda g^{(\epsilon)}_{ab} = 8\pi T_{ab}^{(\epsilon)}
\end{equation}
to associate a 1-parameter family of stress-energy tensors $T_{ab}^{(\epsilon)}$ with the 1-parameter family of metrics $g_{ab}^{(\epsilon)}$. A linearized stress-energy perturbation for this family may be defined by 
\begin{equation}
  \mathcal{T}_{ab} \equiv \left. \partial_\epsilon T_{ab}^{(\epsilon)} \right|_{\epsilon = 0},
\end{equation}
and a standard computation shows that
\begin{align}
  16\pi \mathcal{T}_{ab} = ( \delta^c_a \delta^d_b - \frac{1}{2} \bar{g}_{ab} \bar{g}^{cd} ) [ \bar{\nabla}_f ( 2 \bar{\nabla}_{(c} h_{d)}{}^{f} - \bar{\nabla}^f h_{cd} ) 
  \nonumber
  \\
  ~ -  \bar{\nabla}_c \bar{\nabla}_d h^{f}{}_{f} ] + \bar{g}_{ab} h^{cd} \bar{R}_{cd} + (2 \Lambda - \bar{R}) h_{ab},
  \label{Tcal}
\end{align}
where $\bar{R}_{ab} = R_{ab}^{(0)}$, $\bar{R} = \bar{g}^{ab} \bar{R}_{ab}$, and indices have been raised and lowered using the background metric. For any pair $(\bar{g}_{ab},h_{ab})$, the right-hand side of this expression can be used to find $T_{ab}^{(\epsilon)} = \bar{T}_{ab} + \epsilon \mathcal{T}_{ab} + \ldots$ up to terms which are quadratic and higher order in $\epsilon$.

In the context of this paper, it is more useful to consider the mixed-index stress-energy tensor, the linearization of which we denote by
\begin{align}
  \mathfrak{T}^{b}{}_{a}[h] \equiv \left. \partial_\epsilon (T^{(\epsilon)}_{ac} g^{bc}_{(\epsilon)}) \right|_{\epsilon = 0} = \mathcal{T}_{ac} \bar{g}^{bc} - \bar{T}_{ac} h^{bc} .
\end{align}
The notation on the left-hand side of this equation suggests that $\mathfrak{T}^{b}{}_{a}[h]$ is to be viewed as a linear operator acting on arbitrary rank-2 symmetric tensor fields $h_{ab}$. Using \eqref{Tcal}, it is
\begin{align}
  16\pi \mathfrak{T}^{b}{}_{a} = ( \delta^c_a \bar{g}^{bd} - \frac{1}{2} \delta^b_a \bar{g}^{cd}) [ \bar{\nabla}_f ( 2 \bar{\nabla}_{(c} h_{d)}{}^{f} - \bar{\nabla}^f h_{cd} ) 
  \nonumber
  \\
  -  \bar{\nabla}_c \bar{\nabla}_d h^{f}{}_{f} ] + (\delta^b_a h^{cd} - 2 \delta^c_a h^{bd}) \bar{R}_{cd},
  \label{Tfrak}
\end{align}
which depends on $\Lambda$ only implicitly via $\bar{R}_{ab}$.

One may of course continue to higher orders in perturbation theory by considering more derivatives in $\epsilon$, in which case one would typically add $\epsilon^2 u_{ab} + \epsilon^3 v_{ab} + \ldots$ to the right-hand side of \eqref{gFamily}. It is nevertheless sufficient for our discussion to restrict attention only to first-order perturbations.

\section{Kerr-Schild metrics}
\label{Sect:KS}

Einstein's equation is in general nonlinear, meaning that the linearized mixed-index stress-energy operator $\mathfrak{T}^{b}{}_{a}[h]$ generically describes stress-energy tensors associated with $\bar{g}_{ab} + \epsilon h_{ab}$ only through first order in $\epsilon$. Nevertheless, there exist nontrivial perturbations for which all higher-order corrections vanish identically. One class in which this is known to occur (in several senses) is for Kerr-Schild perturbations \cite{Xanthopoulos1, Xanthopoulos2, Xanthopoulos3, Gursey, Gergely}. A generalized\footnote{The original Kerr-Schild ansatz considered metrics with the form $\eta_{ab} + \epsilon V \ell_a \ell_b$, where $\eta_{ab}$ is flat \cite{KerrSchild}. The metrics which result when replacing $\eta_{ab}$ by a more general background $\bar{g}_{ab}$ are often described as being of generalized Kerr-Schild type \cite{Taub, ExactSolns}. We mostly ignore this distinction.} Kerr-Schild perturbation is one with the form
\begin{equation}
  h^{\mathrm{KS}}_{ab} \equiv V \ell_a \ell_b,
  \label{hKS}
\end{equation}
where $V$ is some scalar and the 1-form $\ell_a$ must be null with respect to the background $\bar{g}_{ab}$. This latter condition implies that $\ell_a$ must also be null with respect to $g_{ab}^{(\epsilon)} = \bar{g}_{ab} + \epsilon h_{ab}^{\mathrm{KS}}$,
\begin{equation}
  g^{ab}_{(\epsilon)} \ell_a \ell_b = \bar{g}^{ab} \ell_a \ell_b = 0.
\end{equation}
Before describing how Kerr-Schild perturbations linearize Einstein's equation, we first remark on their physical and geometrical interpretations.

\begin{figure}
	\includegraphics[width=.55\linewidth]{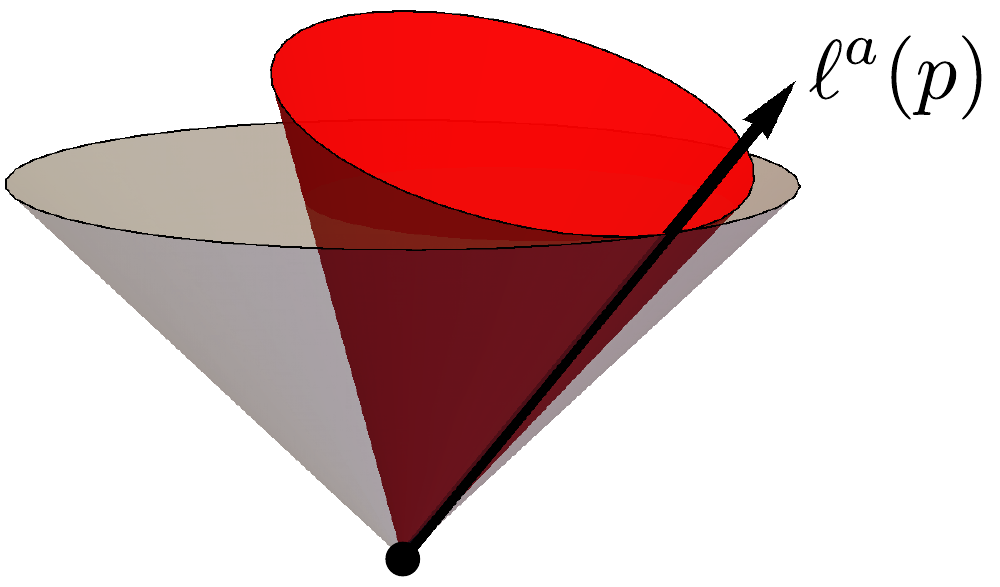}
	\caption{Light cones in a tangent space $T_p M$ associated with $\bar{g}_{ab}$ and $g^{(\epsilon)}_{ab} = \bar{g}_{ab} + \epsilon V \ell_a \ell_b$. They coincide along the ray generated by $\ell^a(p)$, but otherwise differ when $V \ell_a \neq 0$. The cone associated with $\bar{g}_{ab}$ is the inner one when $V(p)<0$ or the outer one when $V(p) > 0$.}
	\label{Fig:Cones}
\end{figure}

Geometrically, Kerr-Schild perturbations deform the light cones of $g_{ab}^{(\epsilon)}$ with respect to those of $\bar{g}_{ab}$. Fixing any event $p$ on the manifold $M$, the two sets of light cones in the tangent space $T_p M$ coincide along the ray proportional to $\ell^a(p) \equiv g^{ab}_{(\epsilon)} \ell_b = \bar{g}^{ab} \ell_b$. Unless $h_{ab}^{ \mathrm{KS} } (p) = 0$, the light cones associated with the two metrics are otherwise distinct; see Fig. \ref{Fig:Cones}. Given an arbitrary vector $v^a \in T_p M$, note that
\begin{equation}
  g_{ab}^{(\epsilon)} v^a v^b = \bar{g}_{ab} v^a v^b + \epsilon V ( \ell_a v^a )^2.
\end{equation}
If $v^a$ is causal with respect to the background (i.e., if $\bar{g}_{ab} v^a v^b \leq 0$), it is guaranteed to remain causal with respect to the perturbed metric whenever $V(p) \leq 0$. Similarly, all vectors which are spacelike with respect to the background remain spacelike whenever $V(p) \geq 0$. Note, however, that these conditions are sufficient, not necessary. 

Another geometric property of Kerr-Schild perturbations is that they identify points in the perturbed and unperturbed spacetimes in such a way that volume elements are exactly preserved:
\begin{equation}
  \epsilon_{abcd} = \bar{\epsilon}_{abcd}.
\end{equation}
Moreover, the inverse $g^{ab}_{(\epsilon)}$ of a Kerr-Schild metric is given by its linearization. Without approximation,
\begin{equation}
  g^{ab}_{(\epsilon)} = \bar{g}^{ab} - \epsilon V \ell^a \ell^b.
\end{equation}
These features follow from the fact that every $h_{ab}^{\mathrm{KS}}$ is trace-free and also a ``square root of zero:''
\begin{equation}
  \bar{g}^{ab} h_{ab}^{\mathrm{KS}} = \bar{g}^{bc} h^{\mathrm{KS}}_{ab} h^{\mathrm{KS}}_{cd} = 0.
  \label{KSprops}
\end{equation}
It is the Lorentzian signature of spacetime which allows for nontrivial 1-forms $\ell_a$ that square to zero, and nontrivial metric perturbations $h_{ab}^{\mathrm{KS}}$ that do the same.

Regardless, once $\bar{g}_{ab}$ is fixed, varying $\ell_a$ and $V$ allows for only three degrees of freedom. It is therefore clear that not all spacetimes can be put into Kerr-Schild form. This may be seen more concretely by noting that under mild assumptions, $\ell^a$ must be a repeated principal null direction for the perturbed metric \cite{ExactSolns}, which can exist only for a small set of spacetimes. Although Kerr-Schild metrics describe only a small class of spacetimes in any particular background, those that they do describe can nevertheless be physically important. If the background is flat, Kerr-Schild perturbations can be used to generate Kerr-Newman black holes, $pp$-waves---which include gravitational plane waves as special cases---Vaidya's collapsing shells, Kinnersley's photon rocket, (anti) de Sitter spacetimes, and more \cite{ExactSolns, dSKerrSchild}. Incidentally, the Kerr-Schild structure has also been used to find ``universal metrics'' which are solutions not only in general relativity, but also in a wide class of modified theories of gravity \cite{Gurses1, Gurses2}.

Without specifying any particular $\bar{g}_{ab}$, we now evaluate the exact Einstein equation for the 1-parameter family of Kerr-Schild metrics $g_{ab}^{(\epsilon)}$. This first requires the difference 
\begin{equation}
  C^{c}{}_{ab} = \frac{1}{2} g^{cd} ( 2 \bar{\nabla}_{(a} g_{b)d} - \bar{\nabla}_d g_{ab} )
\end{equation}
between the connections associated with $\bar{g}_{ab}$ and $g_{ab}^{(\epsilon)}$, which is defined, e.g., so that $\nabla_a \omega_b = \bar{\nabla}_a \omega_b - C^{c}{}_{ab} \omega_c$ for any 1-form $\omega_a$. Applying \eqref{hKS}, one finds that
\begin{equation}
  g_{cd} C^{d}{}_{ab} = \epsilon (\bar{\nabla}_{(a} h^{ \mathrm{KS} }_{b)c} - \frac{1}{2} \bar{\nabla}_c  h^{ \mathrm{KS} }_ {ab}) ,
  \label{connection}
\end{equation}
which is only linear in $\epsilon$. Additionally,
\begin{equation}
  C^{b}{}_{ab} = C^{c}{}_{ab} \ell^a \ell^b = C^{c}{}_{ab} \ell_c \ell^b = 0.
  \label{connectionOrtho}
\end{equation}
These equations imply that the acceleration, expansion, and twist of the null congruence defined by $\ell^a$ are independent of $\epsilon$. Using $\bar{\nabla}_a$ and $\nabla_a$ to denote the covariant derivatives respectively associated with $\bar{g}_{ab}$ and $g_{ab}^{(\epsilon)}$, 
\begin{gather}
  \ell^b \nabla_b \ell_a = \ell^b \bar{\nabla}_b \ell_a , \qquad \nabla_a \ell^a = \bar{\nabla}_a \ell^a,  
  \\
  \nabla_{[a} \ell_{b]} = \bar{\nabla}_{[a} \ell_{b]}.
\end{gather}
The first of these equations implies that if the orbits of $\ell^a$ are (null) geodesics in the background metric, then they are also null geodesics in the deformed metric. The two metrics can therefore share light rays. Indeed, this occurs in all cases we consider in detail; see \eqref{geodesic} below.

Given the deformed connection $C^{c}{}_{ab}$, the deformed Ricci tensor now follows from
\begin{equation}
  R_{ab} = \bar{R}_{ab} + \bar{\nabla}_{c} C^{c}{}_{a b} - C^{d}{}_{bc} C^{c}{}_{ad},
  \label{Ricci}
\end{equation}
where \eqref{connectionOrtho} has been used to eliminate terms involving $C^{b}{}_{ab}$. Substituting \eqref{connection} into this equation results in an expression which is at most cubic in $\epsilon$. Although this is already relatively simple, the cubic component may be eliminated by raising one index. Applying Einstein's equation, the stress-energy tensors of the background and deformed spacetimes are exactly related by
\begin{equation}
	T^{(\epsilon)}_{ac} g^{bc} = \bar{T}_{ac} \bar{g}^{bc} + \epsilon \mathfrak{T}^{b}{}_{a} [h^{\mathrm{KS}}] + \epsilon^2 \mathfrak{S}^{b}{}_{a}[h^{\mathrm{KS}}]
	\label{tKS}
\end{equation}
for Kerr-Schild perturbations, where $\mathfrak{T}^{b}{}_{a}[h]$ is given by \eqref{Tfrak} and
\begin{align}
  8\pi \mathfrak{S}^{b}{}_{a}[h] \equiv ( \delta^b_c \delta_a^d - \frac{1}{2} \delta^b_a \delta^d_c) \big[ \bar{\nabla}_f ( h^{k[f} \bar{\nabla}_k h^{c]}{}_{d} ) 
  \nonumber
  \\
  ~ + \frac{1}{2} \bar{\nabla}^f h^{ck} \bar{\nabla}_d h_{kf} \big].
  \label{frakS}
\end{align}
Note that $\mathfrak{S}^{b}{}_{a}[h]$ returns the correct second-order stress-energy tensor only for perturbations with Kerr-Schild form. The analogous operator for more general perturbations involves additional terms. 

Continuing, \eqref{tKS} implies that Einstein's equation is at worst quadratic in $h^{\mathrm{KS}}_{ab}$ for general Kerr-Schild perturbations with arbitrary matter content. Linearity may be achieved by supposing that 
\begin{equation}
  \mathfrak{T}^{b}{}_{a}[h^{\mathrm{KS}}] \ell^a \ell_b = 0,
  \label{Tll}
\end{equation}
which implies that $V (\ell^a \bar{\nabla}_a \ell_c)( \ell^b \bar{\nabla}_b \ell^c ) = 0$. If $V \neq 0$, it follows that the orbits of $\ell^a$ must be geodesic in the sense that there exists a scalar field $\phi$ such that
\begin{equation}
  \ell^b \bar{\nabla}_b \ell_a = \ell^b \nabla_b \ell_a = \phi \ell_a.
  \label{geodesic}
\end{equation}
Assuming this, a direct calculation using \eqref{frakS} shows that
\begin{equation}
  \mathfrak{S}^{b}{}_{a}[h^{KS}] = 0.
  \label{frakSzero}
\end{equation}
\textit{The mixed-index stress-energy tensor therefore linearizes for Kerr-Schild perturbations constructed from geodesic null congruences}. This result was originally found in flat backgrounds by G\"{u}rses and G\"{u}rsey \cite{Gursey}, and was later generalized by Gergely \cite{Gergely}. From a slightly different logical perspective, Xanthopoulos \cite{Xanthopoulos1} has shown that for (not \textit{a priori} geodesic) vacuum Kerr-Schild deformations of vacuum backgrounds, the geodesic condition is implied and linearity is again achieved. Other variations of this have been discussed as well \cite{Xanthopoulos2, Xanthopoulos3}. In all non-vacuum cases, simple results occur only for the mixed-index form of Einstein's equation. Advantages of this circumstance are explained in Sect. \ref{Sect:Matter} below.

\section{Generating exact solutions from approximate ones}
\label{Sect:genSoln}

Now that we have reviewed the linearization of Einstein's equation which occurs for (geodesic) Kerr-Schild perturbations, it would seem natural to use that linearity to solve Einstein's equation\footnote{A great deal is already known regarding vacuum Kerr-Schild metrics \cite{KerrSchild, ExactSolns, Debney, GergelyCompletesoln}. Nevertheless, our interest here is not in the details of these solutions, but rather in methods which might generalize even for geometries which are not exactly Kerr-Schild.}. As they stand, however, results \eqref{tKS} and \eqref{frakSzero} are considerably less powerful than they might initially appear. They do not imply that, e.g., that Kerr metrics with different origins can be added in a Minkowski background. This is because the sum of two nontrivial Kerr-Schild perturbations $h_{ab}^{\mathrm{KS}} = V \ell_a \ell_b$ and $h'^{\mathrm{KS}}_{ab} = V' \ell'_a \ell'_b$ is not itself Kerr-Schild unless $\ell_a \propto \ell'_a$. The null 1-forms associated with two separated Kerr black holes are not proportional, so linearity results cannot be applied. Nevertheless, there are a few interesting cases in which the relevant 1-forms are proportional. One can, for example, use linearity to explain why there is a sense in which vacuum $pp$-waves propagating in the same direction satisfy exact linear superposition. As a special case, this predicts the known result \cite{Bonnor} that parallel (but not antiparallel) beams of light do not attract. Also, Minkowski perturbations generating de Sitter and Schwarzschild metrics may be added to recover the exact Schwarzschild-de Sitter spacetime. 

These results are rather special. The simplicity associated with the linearized Einstein equation is strongly tempered by the nontrivial algebraic constraints imposed by demanding that metric perturbations be in the Kerr-Schild form \eqref{hKS}. Indeed, the vast majority of detailed calculations in the literature which involve Kerr-Schild metrics have been performed using the rather different set of tools associated with Newman-Penrose or related formalisms. Stated somewhat differently, the ``gauge'' implied by requiring that a metric perturbation be of Kerr-Schild type is not a particularly convenient one. It would be significantly simpler to obtain a linearized perturbation in a gauge which is i) ``unconstrained'' in a reasonable sense, and ii) for which systematic computational methods are known. We now show that in some cases, the errors introduced by working in a convenient gauge can at the end of a calculation be eliminated merely by applying an appropriate gauge transformation.

Consider some particular, not necessarily Kerr-Schild $h_{ab}$ which satisfies the linearized Einstein equation in the sense that
\begin{equation}
  \mathfrak{T}^{b}{}_{a} [h] = \mathfrak{t}^{b}{}_{a}
\end{equation}
for fixed $\mathfrak{t}^{b}{}_{a}$. Also suppose that there exists a gauge vector $\xi^a$ which transforms the metric to Kerr-Schild form:
\begin{equation}
  h_{ab} + \mathcal{L}_\xi \bar{g}_{ab} = h_{ab}^{ \mathrm{KS} } = V \ell_a \ell_b.
  \label{hTrans}
\end{equation}
We now show that the full, nonlinear matter content associated with the gauge-transformed metrics $g_{ab}^{(\epsilon)} = \bar{g}_{ab} + \epsilon h_{ab}^{\mathrm{KS}}$ is essentially identical to the linearized matter content of the (arbitrary-gauge) metrics $\bar{g}_{ab} + \epsilon h_{ab}$. 

Again regarding $\mathfrak{T}^{b}{}_{a}[h]$ as a linear map on symmetric, rank-2 tensor fields, we may consider its behavior under general first-order gauge transformations $h_{ab} \rightarrow h_{ab} + \mathcal{L}_\Xi \bar{g}_{ab}$. Using \eqref{Tfrak}, one finds that for any $\Xi^a$,
\begin{equation}
  \mathfrak{T}^{b}{}_{a} [ h + \mathcal{L}_\Xi \bar{g} ] = \mathfrak{T}^{b}{}_{a} [ h ] + \mathcal{L}_\Xi (\bar{T}_{ac} \bar{g}^{bc}).
  \label{gauge}
\end{equation}
It follows that as usual in perturbation theory \cite{StewartWalker}, $\mathfrak{t}^{b}{}_{a}$ is invariant with respect to all first-order gauge transformations whenever the background stress-energy $\bar{T}_{ac} \bar{g}^{bc}$ has the form $(\mbox{constant}) \times \delta^b_a$. Even if $\mathfrak{t}^{b}{}_{a}$ does depend on gauge, it nevertheless follows from \eqref{tKS}, \eqref{Tll}, \eqref{frakSzero}, and \eqref{gauge} that if
\begin{equation}
  	[\mathfrak{t}^{b}{}_{a} + \mathcal{L}_\xi ( \bar{T}_{ac} \bar{g}^{bc})] \ell^a \ell_b = 0
  	\label{tll}
\end{equation}
for some gauge vector $\xi^a$ which satisfies \eqref{hTrans}, the exact mixed-index stress-energy tensor of the gauge-transformed metric $g_{ab}^{(\epsilon)} = \bar{g}_{ab} + \epsilon V \ell_a \ell_b$ is equal to
\begin{equation}
  T_{ac}^{(\epsilon)} g^{bc}_{(\epsilon)} = [\bar{T}_{ac} \bar{g}^{bc} + \epsilon \mathcal{L}_\xi (\bar{T}_{ac} \bar{g}^{bc})]  + \epsilon \mathfrak{t}^{b}{}_{a} .
  \label{Txform}
\end{equation}
This is our main result. We now consider some of its consequences.

Eq. \eqref{Txform} is most easily interpreted if $h_{ab}$ is a vacuum perturbation to a vacuum background, in which case $\mathfrak{t}^{b}{}_{a} = \bar{T}_{ab} = 0$. All terms on the right-hand side then vanish, implying that the gauge-transformed metric $g_{ab}^{(\epsilon)}$ satisfies the exact vacuum Einstein equation. Even though $\bar{g}_{ab} + \epsilon h_{ab}$ is necessarily vacuum only through first order in $\epsilon$, the errors in this metric exactly cancel those in the first-order approximation $h_{ab} \rightarrow h_{ab} + \mathcal{L}_\xi \bar{g}_{ab}$ to a ``true'' gauge transformation. Sect. \ref{Sect:Examples} provides explicit examples which illustrate how this result can be used to obtain exact vacuum solutions from approximate ones. 

The class of ``well-behaved'' Kerr-Schild spacetimes which are vacuum, asymptotically-flat perturbations on Minkowksi spacetime are unfortunately quite small---essentially just the Kerr solutions \cite{KerrWilson}. A somewhat wider class of interesting applications arise with the consideration of nontrivial matter fields. Suppose then that $\bar{T}_{ab} = 0$, but allow for nontrivial matter content in the perturbation. If $\mathfrak{t}^{b}{}_{a} \ell^a \ell_b = 0$, we then have that
\begin{equation}
 	T_{ac}^{(\epsilon)} g^{bc}_{(\epsilon)} = \epsilon \mathfrak{t}^{b}{}_{a} ,
 	\label{Tcorresp}
\end{equation}
so the mixed-index matter content of the gauge-transformed metrics $\bar{g}_{ab} + \epsilon V \ell_a \ell_b$ is exactly equal to the linearized matter content of the approximate metrics $\bar{g}_{ab} + \epsilon h_{ab}$.

One slight generalization which can sometimes be useful is to assume 
\begin{equation}
  \bar{T}_{ab} = \lambda \bar{g}_{ab}
  \label{Tback}
\end{equation}
for some constant $\lambda$, which may be interpreted as allowing for nontrivial dark energy in the background metric, or alternatively for different cosmological constants in the background and perturbed spacetimes. Regardless, \eqref{tll} and \eqref{Txform} imply that if $\mathfrak{t}^{b}{}_{a} \ell^a \ell_b = 0$,
\begin{equation}
  T_{ac}^{(\epsilon)} g^{bc}_{(\epsilon)} = \lambda \delta^b_a + \epsilon \mathfrak{t}^{b}{}_{a}.
  \label{Tcorresp2}
\end{equation}

In each of these cases, our ability to obtain exact solutions from approximate ones crucially depends on the existence of some vector field $\xi^a$ which solves \eqref{hTrans}. The aforementioned speciality of Kerr-Schild metrics implies that such vector fields cannot exist generically. Nevertheless, there are interesting cases in which appropriate gauge vectors may indeed be found. 


\subsection{Matter properties}
\label{Sect:Matter}

The mixed-index form of results such as \eqref{Tcorresp} provides a remarkably direct interpretation for the matter fields associated with the seed metrics $\bar{g}_{ab} + \epsilon h_{ab}$ and their gauge-transformed analogs: At least in vacuum backgrounds, all eigenvalues and eigenvectors associated with the exact and linearized stress-energy tensors are identical. Somewhat more generally, suppose that \eqref{Tcorresp2} holds and that there exists some eigenvalue $\kappa$ and an associated eigenvector $v^a$ which is associated with the linearized, arbitrary-gauge stress-energy in the sense that
\begin{equation}
  \mathfrak{t}^{b}{}_{a} v^a = \kappa v^b.
\end{equation}
Then
\begin{equation}
  T_{ac}^{(\epsilon)} g^{bc}_{(\epsilon)} v^a = (\lambda + \epsilon \kappa ) v^a,
\end{equation}
implying that the same eigenvector is also associated with the transformed metric. The corresponding eigenvalues are also identical (up to a factor of $\epsilon$) if the background metric is vacuum, i.e. if $\lambda = 0$. Note that in the absence of a unique metric, it does not make sense to compare eigenvectors associated with stress-energy tensors whose indices are both lowered or both raised; these would map vectors into 1-forms, and there is no canonical way to say whether or not such objects might be ``the same.'' It is therefore essential that our result is in a mixed-index form which avoids any need to identify tangent and cotangent spaces.

Results involving the eigenvectors and eigenvalues of a stress-energy tensor directly describe its physical properties. In many cases, eigenvalues can be interpreted as energy densities and principal pressures, while the corresponding eigenvectors can describe the rest frame 4-velocities and the directions associated with the principal pressures. These interpretations must be applied with caution, however, as eigenvectors which are, e.g., timelike in the background may no longer be timelike in the perturbed spacetime. 

It follows from \eqref{Tfrak}, \eqref{Tback}, and $\mathfrak{t}^{b}{}_{a} \ell^a \ell_b = 0$ that the $\ell^a$ associated with the Kerr-Schild structure is always an eigenvector:
\begin{equation}
	( T_{ac}^{(\epsilon)} g^{bc}_{(\epsilon)} ) \ell^a \propto \ell^b.
	\label{lEigen}
\end{equation}
That this eigenvector must be null implies that the matter content of the spacetime, if nontrivial, must be somewhat exotic. 

Another remark which may be made about our ``Kerr-Schild gauge'' is that it provides additional benefits beyond the ability to solve Einstein's equation. One of these is that it can allow for self-consistent discussions of singular matter sources. While it is often useful to work with pointlike or stringlike stress-energy tensors in the linearized theory, doing so in full general relativity is generically meaningless \cite{GerochTraschen}, at least within the context of ordinary linear distribution theory. Nevertheless, there is a sense in which exact distributional stress-energy tensors are known to be consistent within the Kerr-Schild class of metrics \cite{VickersColombeau}. This is essentially a corollary of the linearity result: Suppose again that $h_{ab}$ is some given perturbation---not necessarily in Kerr-Schild form---and that $\mathfrak{t}^{b}{}_{a} = \mathfrak{T}^{b}{}_{a}[h]$ is distributional. Also suppose that \eqref{tll} holds, and that there exists a gauge vector $\xi^a$ which solves \eqref{hTrans}. The remapped metric may then be \textit{assigned} an exact distributional stress-energy tensor via \eqref{Txform}. Note however that this provides a well-defined distribution for $T_{ac}^{(\epsilon)} g^{bc}_{(\epsilon)}$, treated as one object, but not to $T_{ac}^{(\epsilon)}$ alone.

As an example of this, consider the exact Schwarzschild metrics with masses $\epsilon m$ as Kerr-Schild perturbations $g_{ab}^{(\epsilon)} = \eta_{ab} + \epsilon V \ell_a \ell_b $ to the flat metric $\eta_{ab}$. In appropriate Minkowski coordinates $x^\mu = (t,x^i)$, the perturbation is explicitly
\begin{equation}
	V = \frac{2m}{r}, \qquad \ell_\mu d x^\mu = - ( dt +  \hat{\bm{n}} \cdot d\bm{x}  ),
	\label{Schw}
\end{equation}
where $\hat{ \bm{n} } \equiv \bm{x}/r$ and $r \equiv |\bm{x}| = \sqrt{ \delta_{ij} x^i x^j }$. A direct calculation shows that this perturbation is sourced, without approximation, by the ``expected'' point mass expression \cite{distSchw}
\begin{equation}
  T_{\mu\rho}^{(\epsilon)} g^{\nu\rho}_{(\epsilon)} = - \epsilon m \delta^3(\bm{x}) \delta^t_\mu \delta^\nu_t.
  \label{Tschw}
\end{equation}
Distributional stress-energy tensors have in fact been computed for generic members of the Kerr (-Newman) family of spacetimes \cite{McManus, Balasin}, and these in general involve singular sources which are extended with respect to the flat background structure.

\subsection{Conservation laws}

Another benefit of the Kerr-Schild structure is that it provides particularly simple relations between conservation laws in the full spacetime and conservation laws in the background.

We first note that as an integrability condition on the full Einstein equation, the Bianchi identity implies that $\nabla_b (T_{ac}^{(\epsilon)} g^{bc}_{(\epsilon)}) = 0$, at least for nonsingular stress-energy tensors. Rewriting this in terms of the background derivative operator using \eqref{connection},
\begin{align}
  \bar{\nabla}_b (T_{ac}^{(\epsilon)} g^{bc}_{(\epsilon)}) = \frac{1}{2} \epsilon (T^{(\epsilon)}_{df} \bar{g}^{bd} \bar{g}^{cf} ) \bar{\nabla}_a h_{bc}^{\mathrm{KS}}.
\end{align}
If $\ell^a$ is an eigenvector of the stress-energy tensor, as is implied by the assumptions stated immediately above \eqref{lEigen}, the right-hand side here vanishes and 
\begin{equation}
	\bar{\nabla}_b (T_{ac}^{(\epsilon)} g^{bc}_{(\epsilon)}) = 0.
	\label{backTcons}
\end{equation}
It follows that at least when the background metric satisfies \eqref{Tback} and when $\mathfrak{t}^{b}{}_{a} \ell^a \ell_b = 0$, stress-energy conservation holds with respect to both the deformed and background metrics. This result is easily extended to apply also for distributional stress-energy tensors by directly computing from \eqref{Tfrak} that $\bar{\nabla}_b \mathfrak{T}^{b}{}_{a}[h] = 0$ whenever $\bar{R}_{ab} = (\mathrm{constant}) \times \bar{g}_{ab}$. Similarly, \textit{test matter} with stress-energy $t_{ab} = t_{(ab)}$  which satisfies $\nabla_b (t_{ac} g^{bc}_{(\epsilon)} ) = 0$ and for which $\ell^a$ is an eigenvector also satisfy $\bar{\nabla}_b (t_{ac} g^{bc}_{(\epsilon)} ) = 0$.

It is well-known that if there exists a Killing field associated with some metric, stress-energy conservation with respect to that metric implies the existence of a conserved current which is linear in the stress-energy tensor. This result is not necessarily very useful, as there may not be many Killing fields in the deformed spacetime. Eq. \eqref{backTcons} nevertheless implies that there are cases in which \textit{background} Killing fields may also be used to generate conserved currents. We first note from \eqref{Tfrak} that if the background stress-energy satisfies \eqref{Tback}, $\bar{g}^{a[b} \mathfrak{T}^{c]}{}_{a}[h]  = 0$ for all $h_{ab}$. Additionally assuming that $\mathfrak{t}^{b}{}_{a} \ell^a \ell_b = 0$, it follows from \eqref{Tcorresp2} that
\begin{equation}
	(T_{cd}^{(\epsilon)} g^{c[a}_{(\epsilon)}) \bar{g}^{b]d} = 0.
\end{equation}
The current
\begin{equation}
  j^a_\psi \equiv (T_{bc}^{(\epsilon)} g^{ca}_{(\epsilon)}) \psi^b 
\end{equation}
is therefore conserved for every $\psi^a$ satisfying $\mathcal{L}_\psi \bar{g}_{ab} = 0$. Moreover, it is conserved with respect to the background and perturbed metrics:
\begin{equation}
   \nabla_a j^a_\psi = \bar{\nabla}_a j^a_\psi = 0.
\end{equation}
If the background is flat---or more generally maximally-symmetric---this provides ten conservation laws for the stress-energy associated with a Kerr-Schild spacetime. It is clear for example that the Schwarzschild stress-energy \eqref{Tschw} satisfies all ten special-relativistic conservation laws; it has the interpretation of a point particle at rest in the flat background. Similar conservation laws also hold for test stress-energy tensors on Kerr-Schild spacetimes for which $\ell^a$ is an eigenvector.

\section{Examples}
\label{Sect:Examples}

We now describe some simple examples which illustrate how gauge transformations can be used to generate exact solutions from approximations. All examples here use a flat background, $\bar{g}_{ab} = \eta_{ab}$, and we shall often use coordinates $(t,x^i)$ which are Minkowski with respect to this background. 

\subsection{Spherical symmetry}
\label{Sect:Sph}

One standard approximation for a (not necessarily spherical) metric is the post-Newtonian expansion, and the lowest-order version of this---the Newtonian metric---can be written as \cite{PoissonWill, Wald}
\begin{equation}
  h_{ab} = - 2 \phi (\eta_{ab} + 2 \nabla_a t \nabla_b t)
  \label{hNewton}
\end{equation}
in terms of the Newtonian potential $\phi$. If time derivatives are neglected, this satisfies the Lorenz gauge condition
\begin{equation}
  \bar{\nabla}^b ( h_{ab} - \frac{1}{2} \eta_{ab} h^{c}{}_{c}) = 0.
\end{equation}
 
We now restrict for simplicity to spherical, time-independent Newtonian perturbations, in which case $\phi = \phi(r)$ for an appropriate radial coordinate $r = |\bm{x}|$. Our discussion in Sect. \ref{Sect:genSoln} implies that the corresponding approximation can be improved if there exists a gauge vector $\xi^a$ which transforms $h_{ab}$ into Kerr-Schild form. More specifically, we seek some triple $(\xi^a, V, \ell_a)$ which is a solution to \eqref{hTrans}. It suffices here to consider the radially-ingoing null 1-form in \eqref{Schw}, a static, spherically-symmetric Kerr-Schild potential $V = V(r)$, and a gauge vector with the form $\xi^\mu \partial_\mu = \xi^0(r) \partial_t + \xi^r(r) \partial_r$. The various components of the gauge transformation equation then imply the equalities
\begin{align}
  V = -2 \phi = - \frac{d\xi^0}{dr} = - 2 \xi^r/r = 2 \left( \frac{d\xi^r}{dr} - \frac{\xi^r}{r} \right).
\end{align}
Eliminating $\xi^r$ from these equations in favor of $V$, it follows as a necessary condition for the existence of an appropriate gauge transformation that the Kerr-Schild potential must satisfy
\begin{align}
  \frac{dV}{dr} + V/r = 0,
  \label{Vsph}
\end{align}
the only solutions of which are
\begin{equation}
  V = - 2 \phi  = \frac{ 2 m }{ r }
  \label{Vschw}
\end{equation}
for some constant $m$. We have thus recovered the exact Schwarzschild solution \eqref{Schw} from the Newtonian, harmonic-gauge approximation to the metric sourced by a point particle\footnote{The Newtonian mass density associated with this $\phi$ is $\rho_N = (4\pi)^{-1} \bar{\nabla}^2 \phi = m \delta^3 (\bm{x})$. The Newtonian mass in the seed potential is exactly equal to the ADM mass in the gauge-transformed metric.} with mass $m$. It can be generated by the explicit gauge vector
\begin{equation}
  \xi^\mu \partial_\mu = - m ( \ln r^2 \partial_t + \hat{n}^i \partial_i).
  \label{gaugeSchw}
\end{equation}
Note that even though $h_{ab}$ and $h_{ab}^{\mathrm{KS}}$ both decay at large distances, $\xi^a$ does not; indeed it grows logarithmically. 

It is also worth remarking that the constraint \eqref{Vsph} on $V$ was not obtained from Einstein's equation. It appears instead as a consistency condition required by the ability to transform spherical, Lorenz-gauge Newtonian perturbations into Kerr-Schild perturbations. As such, the \textit{only} static, spherically-symmetric Newtonian spacetime for which our result applies, with a flat background, is Schwarzschild.

Some (rather exotic) ``stellar interiors'' may be obtained by considering static, spherically-symmetric perturbations with the form
\begin{align}
  h_{ab} = \psi (\eta_{ab} + 2 \nabla_a r \nabla_b r) - 2 \phi_N (\eta_{ab} + 2 \nabla_a t \nabla_b t) 
  \nonumber
  \\
  ~ + \zeta \eta_{ab},
\end{align}
where $\phi$, $\psi$, and $\zeta$ are all assumed to depend only on $r$. This satisfies the Lorenz gauge condition whenever
\begin{equation}
 	\frac{d\zeta}{dr } = 4 \psi/r.
 	\label{LorenzSph}
\end{equation}
For any potentials, the linearized stress-energy $\mathfrak{t}^{b}{}_{a}= \mathfrak{T}^{b}{}_{a}[h]$ admits four eigenvectors corresponding to the spherical coordinate directions $\partial_t$, $\partial_r$, $\partial_\theta$, and $\partial_\varphi$. The principal pressures $p_\perp$ associated with the two angular directions are equal and are given by
\begin{equation}
  p_\perp = \frac{1}{8\pi} \left(  \frac{d^2 \psi}{dr^2} + \frac{4}{r} \frac{d\psi}{dr} \right),
\end{equation}
while the eigenvalues associated with the temporal and radial directions are in general different. If $h_{ab}$ can be transformed into Kerr-Schild form, it follows from \eqref{Tcorresp} and  \eqref{lEigen} that $\ell^a$ must be an eigenvector for $\mathfrak{t}^{b}{}_{a}$. It is therefore necessary that we consider potentials in which the temporal and radial eigenvalues  coincide with one another (but not necessarily with $p_\perp$).

This may be seen by explicitly writing down the various relations implied by the gauge transformation equation \eqref{hTrans}. One consequence of that equation is
\begin{equation}
	\frac{dV}{dr} + V/r = -\frac{2}{r^3} \frac{d}{dr} (r^3 \psi),
	\label{Vsph2}
\end{equation}
which generalizes \eqref{Vsph}. For fixed $\psi$, this determines the Kerr-Schild potential up to the addition of a Schwarzschild-like term proportional to $1/r$. The constant is fixed (when possible) by an additional consequence of \eqref{hTrans}, namely
\begin{equation}
  V = - 2 \phi - (\psi + \zeta).
\end{equation}
The combination of this equation with \eqref{Vsph2} may also be viewed as a consistency relation on the allowable seed perturbations $h_{ab}$. It implies, for example, that the potentials must satisfy
\begin{equation}
  \bar{\nabla}^2 \phi = 4\pi \rho_N =\frac{1}{2} \left( \frac{d^2 \psi}{dr^2} + \frac{2}{r} \frac{d\psi}{dr} - 4 \psi/r^2 \right).
\end{equation}
The radial and temporal eigenvalues of $\mathfrak{t}^{b}{}_{a}$ then coincide, as claimed, and are equal to
\begin{equation}
	p_{\|} = - \rho = \frac{1}{4\pi r^4} \frac{d}{dr} (r^3 \psi).
	\label{pPar}
\end{equation}
Reintroducing the constant $\epsilon$, the transverse and longitudinal pressures $\epsilon p_\perp$, $\epsilon p_\|$, as well as the energy density $\epsilon \rho$, are the \textit{exact} eigenvalues for the stress-energy tensor associated with the gauge-transformed Kerr-Schild metrics $g_{ab}^{(\epsilon)} = \eta_{ab} + \epsilon V \ell_a \ell_b$. 

In the vacuum case where $p_\| = p_\perp = 0$, the most general solution for $\psi$ must be proportional to $1/r^3$, but it is clear from \eqref{Vsph2} that solutions of this type can only generate Schwarzschild solutions; they are no more general than seed perturbations for which $\psi = 0$. In non-vacuum cases, \eqref{Vsph2} and \eqref{pPar} may be combined to directly relate the Kerr-Schild potential $V$ to the physical parameter $p_\|$. If the metric is Schwarzschild with mass $m$ for all $r > r_\mathrm{surf}$,
\begin{equation}
  V(r) = \frac{2}{r} \left( m - 4\pi \int_r^{r_\mathrm{surf}} \rho (\mathfrak{r}) \mathfrak{r}^2 d \mathfrak{r} \right).
\end{equation}
Even in a non-vacuum region, $V$ therefore has the form $2 \times (\mbox{enclosed mass})/r$. 

Unfortunately, even these solutions are not particularly general. It is however known that all static, spherically-symmetric metrics can be put into Kerr-Schild form with a background which is \textit{conformally} flat \cite{sphKS}. One might therefore attempt to generalize our above results by still setting $\bar{g}_{ab} = \eta_{ab}$, but now allowing for an additional, conformal degree of freedom: Suppose that there exists some gauge vector $\xi^a$ satisfying
\begin{equation}
  h_{ab} + \mathcal{L}_\xi \eta_{ab} = \Omega \eta_{ab} + V \ell_a \ell_b,
\end{equation}
which reduces to \eqref{hTrans} when $\omega = 0$, and corresponds to the mapping
\begin{equation}
  \eta_{ab} + \epsilon h_{ab} \mapsto g_{ab}^{(\epsilon)}= (1+ \epsilon \Omega) \eta_{ab} + \epsilon V \ell_a \ell_b. 
\end{equation}
Returning for simplicity to the Newtonian perturbations \eqref{hNewton}, the transformed metric may be shown to be determined by
\begin{align}
  \frac{dV}{dr} - V/r = -4 \frac{d\phi}{dr}, \qquad   \Omega = V + 2 \phi
  \label{Vconf}
\end{align}
in terms of the seed potential $\phi$. The correspondence rules between perturbed and gauge-transformed metrics which are derived in Sect. \ref{Sect:genSoln} do not hold with this additional conformal degree of freedom, so we cannot immediately conclude anything about the matter content of the transformed spacetime. It is nevertheless interesting to investigate this directly.

The first point which may be noted is that the transformed metric is not unique for a given perturbation. The integration constant which arises when solving \eqref{Vconf} instead returns a two-parameter family of transformed metrics (one parameter being the perturbation amplitude $\epsilon$). Even a point-particle $\phi = -m/r$ seed results in $V = 2m/r + \gamma r$ and $\Omega = \gamma r$ for any constant $\gamma$, which corresponds to the exact Schwarzschild solution only if $\gamma =0$. We therefore see that, e.g., seed metrics which are vacuum through linear order do not necessarily map into metrics which are vacuum to all orders. Nevertheless, the eigenvalues of the linearized and exact stress-energy tensors \textit{are} related. For any $\phi$,
\begin{equation}
  (1 + \epsilon \Omega)^2 \left[ \rho_{(\epsilon)} + \sum_i p^i_{(\epsilon)} \right] = \epsilon \rho_N,
  \label{confKS}
\end{equation}
where $\rho_{(\epsilon)}$ is the energy density associated with $T_{ac}^{(\epsilon)} g^{bc}_{(\epsilon)}$, the $p^i_{(\epsilon)}$ are its three principal pressures, and $\rho_N$ is the Newtonian mass density associated with $\phi$. The factor of $(1+ \epsilon \Omega)^2$ on the left-hand side corresponds to the proportionality factor
\begin{equation}
  \epsilon_{abcd} =  (1 + \epsilon \Omega)^2 \bar{\epsilon}_{abcd}
\end{equation}
between the volume elements associated with the deformed and background metrics. Its presence in \eqref{confKS} suggests that the seed and transformed metrics have identical \textit{volume-weighted} notions of $\rho + \sum_i p^i$. This was also remarked upon in \cite{HarteNonlinear}, although no details were given there. We present it here as a curiosity which might provide some insight into generalizing the (better justified but more specialized) non-conformal Kerr-Schild procedure discussed above. It suggests that weakening some assumptions may result in \textit{parts of} Einstein's equation still being solved exactly by appropriate transformations of convenient approximations. Imposing selection criteria on the relevant integration constants may improve this correspondence even further.

As one final comment on generalizations, it can sometimes be useful to also incorporate information from higher-order perturbation theory. Consider for example a family of spherically-symmetric bodies with masses $\epsilon m$ and electric charges $\epsilon q$. The stress-energy tensors associated with these systems are quadratic in the electromagnetic field, and therefore quadratic in $\epsilon$. It follows that first-order perturbations to flat spacetime must be independent of $q$, implying that we may adopt the uncharged point-mass perturbation $h_{ab}$ which is given by combining \eqref{hNewton} and \eqref{Vschw}. Nonzero charge can be incorporated by also including the second-order metric perturbation
\begin{equation}
	u_{ab} = \frac{1}{2r^2} [ (3m^2-q^2) \eta_{ab} - (m^2+3q^2) \nabla_a t \nabla_b t].
\end{equation}
As we have already discussed, applying the first-order gauge transformation $\mathcal{L}_\xi \eta_{ab}$ which is generated by \eqref{gaugeSchw} transforms $h_{ab}$ into the Kerr-Schild perturbation associated with the exact Schwarzschild metric. Now consider instead a second-order gauge transformation which places both $h_{ab}$ and $u_{ab}$ into Kerr-Schild form. This requires a second-order gauge vector $\chi^a$ which satisfies
\begin{equation}
	u_{ab} + \mathcal{L}_\chi \eta_{ab} + \mathcal{L}_\xi (h_{ab} + \frac{1}{2} \mathcal{L}_\xi \eta_{ab})  = U \ell_a \ell_b
\end{equation}
for some $U$, where $\ell_a$ is the radially-ingoing null 1-form given in \eqref{Schw}. All solutions require that $U = -q^2/r^2$, and one possible gauge vector is explicitly
\begin{equation}
	\chi^\mu \partial_\mu = \frac{1}{r} [ (3m^2-q^2) \partial_t - \frac{1}{4} (m^2-q^2) \hat{n}^i \partial_i ].
\end{equation}
Applying the corresponding gauge transformation to the approximate metrics $\eta_{ab} + \epsilon h_{ab} + \epsilon^2 u_{ab}$ now recovers the \textit{exact} Reissner-Nordstr\"{o}m metrics with masses $\epsilon m$ and charges $\epsilon q$:
\begin{align}
	g_{ab}^{(\epsilon)} &= \eta_{ab} + \epsilon ( h_{ab} + \mathcal{L}_\xi \eta_{ab}) + \epsilon^2 [u_{ab}  + \mathcal{L}_\chi \eta_{ab} 
	\nonumber
	\\
	& \qquad \qquad \qquad \qquad ~ + \mathcal{L}_\xi ( h_{ab} + \frac{1}{2} \mathcal{L}_\xi \eta_{ab})]
	\nonumber 
	\\
	&= \eta_{ab} + \left( \frac{2\epsilon m}{r} - \frac{ \epsilon^2 q^2 }{ r^2 } \right) \ell_a \ell_b.
\end{align}
We make no further attempt to discuss higher-order perturbations in this paper.

\subsection{Axisymmetry and Kerr}
\label{Sect:Kerr}

We now derive the exact Kerr solution from its first-order post-Minkowskian approximation. It is necessary that one be somewhat precise about what this means: One cannot, for example, consider the ``Newtonian limit'' of Kerr---that would return the Schwarzschild result discussed above. Neither can one obtain the desired result by adding the linear-in-spin correction to the Newtonian approximation. The multipole structure of a linearized solution is gauge-invariant, and so the relevant seed solution must incorporate the full, infinite set of nonvanishing multipole moments associated with the exact Kerr solution. We must therefore start with the first post-Minkowski (and not post-Newtonian) approximation.

Assuming $\Lambda =0$, all asymptotically-flat solutions outside of compact sources are known for the vacuum Einstein equation linearized about Minkowski spacetime \cite{BlanchetLRR, BlanchetDamour, Thorne}. Specializing to a regime which is stationary with respect to some Minkowksi time coordinate $t$, these solutions can be written as\footnote{The current moments $J_L$ which appear here are normalized to match Hansen's definition \cite{Hansen}, which differs by a factor of $2l/(l+1)$ \cite{MomentEq} from the more common normalizations used by Thorne \cite{Thorne} and by Blanchet and Damour \cite{BlanchetDamour}.}
\begin{align}
  h_{ab}& = \sum_{l=0}^\infty \frac{(-1)^l}{l!} \bar{\nabla}_L \left( \frac{ 2 I^L }{ r } \right) ( \eta_{ab} + 2 \bar{\nabla}_a t \bar{\nabla}_b t)
  \nonumber
  \\
  & ~ + 4 \bar{\epsilon}_{(a}{}^{cdf} \bar{\nabla}_{b)} t \bar{\nabla}_c t \sum_{l=1}^\infty \frac{(-1)^l}{l!} \bar{\nabla}_d \bar{\nabla}_{L-1} \left( \frac{J_{f}{}^{L-1}}{r} \right) 
  \label{hBlanchet}
\end{align}
up to first-order gauge transformations, where $r$ is an ordinary radial coordinate and $L$ is a multi-index of length $l$ (so, e.g., $\bar{\nabla}_L = \bar{\nabla}_{a_1} \cdots \bar{\nabla}_{a_l}$). The rank-$l$ tensors $I^L$ and $J^L$ which appear here are the mass and current multipole moments, respectively, all of which are symmetric, trace-free, and spatial. If the perturbation is axisymmetric about an axis parallel to a constant unit vector $\hat{z}^a$, there exist scalar coefficients $I_l$ and  $J_l$ such that
\begin{equation}
  I^L = I_l \hat{z}^{\langle L \rangle}, \qquad J^L = J_l \hat{z}^{\langle L \rangle},
  \label{momentsAxisym}
\end{equation}
where $\hat{z}^{\langle L \rangle}$ denotes the symmetric, trace-free component of $\hat{z}^{a_1} \cdots \hat{z}^{a_l}$. The spherical Newtonian limit described by \eqref{hNewton} and \eqref{Vschw} corresponds in this context to setting all moments to zero except for $I_0 = m$.

We now generalize this by considering a 2-parameter family of linearized, axisymmetric perturbations whose moments satisfy
\begin{equation}
  I_l + i J_l = m (i a)^l.
  \label{KerrMoments}
\end{equation}
in terms of the mass\footnote{This is really a normalized mass parameter. The perturbation $\eta_{ab} + \epsilon h_{ab}$ has linearized mass $\epsilon m$ and linearized angular momentum $\epsilon m a$.} $m = I_0$ and the angular momentum per unit mass $a = J_1/m$. They are the multipole moments known to characterize the exact Kerr solution \cite{Hansen}. Using them, the two series in \eqref{hBlanchet} may be summed explicitly.

Combining \eqref{hBlanchet}, \eqref{momentsAxisym}, and \eqref{KerrMoments} while defining $\bm{a} \equiv a \hat{\bm{z}}$, the series in $h_{ab}$ involving the mass moments can be reduced to 
\begin{align}
   h_{00} &= \cos (\bm{a} \cdot \bar{\bm{\nabla}}) \left( \frac{ 2m }{ r } \right)
   \nonumber
   \\
	&= m \left( \frac{1}{ | \bm{x} + i \bm{a} | } + \frac{1}{ | \bm{x} - i \bm{a} | } \right),
	\label{h00}
\end{align}
where the second equality has used that the cosine may be split into a pair of complex exponentials $\exp(\pm i \bm{a} \cdot \bar{\bm{\nabla}})$, each of which acts as a translation operator on $2m/r$. The result may be interpreted as the (real) Newtonian potential in between two point masses which have an imaginary offset between them.

We now turn to the second series in \eqref{hBlanchet}, involving the current moments $J_L$. Again using \eqref{momentsAxisym} and \eqref{KerrMoments}, it may be written as
\begin{align}
  h_{0i} &= -\bar{\epsilon}_{i}{}^{jk} \bar{\nabla}_j \left[ a_k \sinc (\bm{a} \cdot \bar{\bm{\nabla}}) \left( \frac{2m}{r} \right) \right],
  \label{h0iInit}
\end{align}
where $\sinc y \equiv \sin y/y$ and $\bar{\epsilon}_{ijk} \equiv \bar{\epsilon}_{0ijk}$. The right-hand side is implicitly still an infinite series, although it may be summed by noting that
\begin{equation}
	(\bm{a} \cdot \bar{\bm{\nabla}}) \ln (r \pm z) = \pm \frac{a}{r},
\end{equation}
in terms of $z \equiv \hat{\bm{z}} \cdot \bm{x}$. Substituting this identity into \eqref{h0iInit} results in an expression involving $[\sinc (\bm{a} \cdot \bar{\bm{\nabla}}) ] (\bm{a} \cdot \bar{\bm{\nabla}})  = \sin (\bm{a} \cdot \bar{\bm{\nabla}})$, which may in turn be expressed as a difference between two exponential operators. Effecting the associated translations finally results in
\begin{align}
  h_{0i} &= m\bar{\epsilon}_{i}{}^{jk} \hat{z}_j \bar{\nabla}_k \left[ \sin (\bm{a} \cdot \bar{\bm{\nabla}} ) \ln \left( \frac{ r + z }{ r - z } \right) \right]
  \nonumber
  \\
  &= im \left( \frac{ \bar{\epsilon}_{ijk} x^j \hat{z}^k }{ r^2-z^2 } \right) \left[ \frac{ z - i a}{ | \bm{x} - i \bm{a} |} - \frac{ z + i a}{ | \bm{x} + i \bm{a} |} \right].
  \label{h0i}
\end{align}

This can be made more intuitive by introducing coordinates which are better adapted to the system. It is clear from the discussion thus far that complex translations play an important role, which suggests that it may be useful to define the complex-translated radial coordinate
\begin{equation}
  \tilde{r} \equiv \cos(\bm{a} \cdot \bar{\bm{\nabla}}) r = \frac{1}{2} ( | \bm{x} + i \bm{a} | + | \bm{x} - i \bm{a} | ).
  \label{rObl}
\end{equation}
It is also useful to define an angular coordinate $\tilde{\theta}$ via
\begin{equation}
	\cos \tilde{\theta} \equiv z/\tilde{r},
	\label{thetaObl}
\end{equation}
which reduces to the ordinary polar angle when $a = 0$. More generally, $\tilde{r}$ and $\tilde{\theta}$ are known as oblate spheroidal coordinates. In terms of them, 
\begin{equation}
  |\bm{x} \pm i \bm{a}| = \tilde{r} \pm i a \cos \tilde{\theta}, 
\end{equation}
which allows \eqref{hBlanchet}, \eqref{h00}, and \eqref{h0i} to be summarized by
\begin{align}
  h_{ab} = \frac{2m \tilde{r} }{ \tilde{r}^2 + a^2 \cos^2 \tilde{\theta} } \bigg[ ( \eta_{ab} + 2 \bar{\nabla}_a t \bar{\nabla}_b t)
  \nonumber
  \\
  ~ +   \frac{ 2 x_d a_f \bar{\epsilon}_{(a}{}^{cdf} \bar{\nabla}_{b)} t \bar{\nabla}_c t }{ \tilde{r}^2 + a^2 }  \bigg].
  \label{hKerr}
\end{align}
The metrics $\eta_{ab} + \epsilon h_{ab}$ describe linearized, first post-Minkowskian, Lorenz-gauge metrics with the same multipole structure as exact Kerr solutions having masses $\epsilon m$ and angular momenta $\epsilon m a$. We emphasize however that these metrics are not exact solutions to Einstein's equation; their Ricci tensors are nonzero and of order $\epsilon^2$.

Our next step is to use this approximation to derive the exact Kerr metric. The discussion in Sect. \ref{Sect:genSoln} guarantees that this is possible if there exists a gauge vector $\xi^a$ which satisfies \eqref{hTrans} for some Kerr-Schild potential $V$ and some null 1-form $\ell_a$. There is no loss of generality in demanding that the time component of this 1-form is normalized such that
\begin{equation}
  \ell^a \bar{\nabla}_a t = 1,
  \label{ellT}
\end{equation}
in which case the time-time component of the gauge transformation equation immediately shows that 
\begin{equation}
  V = \cos (\bm{a} \cdot \bar{\bm{\nabla}}) \left( \frac{ 2m }{ r } \right) = \frac{ 2 m \tilde{r} }{ \tilde{r}^2 + a^2 \cos^2 \tilde{\theta} } .
  \label{V}
\end{equation}
The second equality here re-expresses \eqref{h00} in terms of the oblate spheroidal coordinates defined by \eqref{rObl} and \eqref{thetaObl}. 

If an appropriate gauge vector exists and \eqref{ellT} is assumed, the generated Kerr-Schild potential must be equal to \eqref{V}. Our next step is to find the spatial components $\ell_i$ of the Kerr-Schild 1-form. Eq. \eqref{h0i} and the time-space components of \eqref{hTrans} show that this must satisfy
\begin{align}
  V \ell_i   &= V \left( \frac{ \bar{\epsilon}_{ijk} a^j x^k }{ \tilde{r}^2+a^2 } \right) + \bar{\nabla}_i \xi^0,
  \label{Vell}
\end{align}
which acts as a Helmholtz decomposition for $V \ell_i$. If we assume that $\xi^0$ is axisymmetric, the norm of this equation may be used together with the requirement that $\ell_i \ell^i = 1$ to deduce both $\xi^0$ and $\ell_i$. Evaluating that norm and noting from \eqref{rObl} that
\begin{equation}
  | \bar{\bm{\nabla}} \tilde{r} |^2 = \frac{ \tilde{r}^2 + a^2 }{ \tilde{r}^2 + a^2 \cos^2 \tilde{\theta} } ,
\end{equation}
one finds that it is sufficient to suppose that $\xi^0$ depends only on $\tilde{r}$. The condition that $\ell_i$ have unit length then reduces to the ordinary differential equation
\begin{align}
	\frac{ d \xi^0 }{ d\tilde{r} } = - \frac{ 2 m \tilde{r} }{ \tilde{r}^2 + a^2 },
\end{align}
which is easily solved to yield
\begin{equation}
  \xi^0 = -m \ln (\tilde{r}^2 + a^2).
\end{equation}
Substituting this back into \eqref{Vell} finally shows that
\begin{equation}
  \ell_i = \frac{ 1 }{ \tilde{r}^2+a^2 } \left[ \bar{\epsilon}_{ijk} a^j x^k  - (\tilde{r}^2 + a^2 \cos^2 \tilde{\theta} ) \bar{\nabla}_i \tilde{r} \right].
  \label{li}
\end{equation}

Some geometrical interpretation for this vector field may be gained by differentiating, 
\begin{equation}
\bar\nabla_i\ell_j=\frac{-\tilde r(\delta_{ij}-\ell_i\ell_j)+a\cos\tilde\theta\, \bar\epsilon_{ijk}\ell^k}{\tilde r^2+a^2\cos^2\tilde\theta},
\label{Dl}
\end{equation}
from which it follows that $\ell^i$ is geodesic and shear-free on the Euclidean $t = \mathrm{const}.$ hypersurfaces.  It is tangent to a congruence of straight lines lying within the one-sheeted hyperboloids $\tilde\theta=\textrm{constant}$.  The lines form kinks across the disk defined by $z=0$ and $r \leq a$, where they meet their $z \rightarrow -z$ mirror images; this arises from a reversal between the two possible ``winding directions'' for straight lines embedded into hyperboloids.  The congruence is smooth everywhere off of the disk.

Together with \eqref{ellT} and \eqref{V}, the $\ell_i$ given by \eqref{li} completely describes the Kerr-Schild metric perturbation which must arise if an appropriate gauge transformation is possible. The last remaining step to showing that such a gauge transformation is indeed possible is to find some $\xi_i$ which satisfies the purely spatial components of \eqref{hTrans},
\begin{align}
	V ( \delta_{ij} - \ell_i \ell_j ) + 2 \bar{\nabla}_{(i} \xi_{j)} = 0.
	\label{gaugeSpaceSpace}
\end{align}
Comparison with \eqref{Dl} immediately reveals that $\xi_i=m \ell_i$ is a solution.  Thus, a full gauge vector which transforms the Lorenz-gauge $h_{ab}$ into Kerr-Schild form is
\begin{equation}
  \xi^\mu \partial_\mu = - m [ \ln (\tilde{r}^2 + a^2) \partial_t - \ell^i \partial_i ].
  \label{gaugeKerr}
\end{equation}
Applying this to the approximate metrics $\eta_{ab} + \epsilon h_{ab}$ results in $g_{ab}^{(\epsilon)} = \eta_{ab} + \epsilon V \ell_a \ell_b$, where $V$ is explicitly given by \eqref{V}, and $\ell_a$ by \eqref{ellT} and \eqref{li}. These are exact Kerr metrics, in Kerr-Schild form, with masses $\epsilon m$ and angular momenta $\epsilon m a$. While there are now many derivations of the Kerr metric \cite{KerrSchild, JanisNewman, KerrBini, SymKerr}, ours has a rather different character from the others: It is much more closely connected to the general-purpose tools commonly used in perturbation theory.

Geometrically, the orbits of the Kerr-Schild $\ell^a$ are null geodesics with respect to both the background and Kerr geometries, so ``background light rays'' which fall into a Kerr black hole with the appropriate $a$-dependent ``offset'' compatible with \eqref{ellT} and \eqref{li} are also physical light rays which can exist in Kerr; they are unaffected by the geometry change. It is also interesting to note that the Kerr-Schild quantities are almost identical to those associated with the Schwarzschild solution characterized by \eqref{Schw}. The Kerr potential $V$ can, e.g., be obtained via a pair of complex translations acting on the Schwarzschild potential. Similar comments regarding complex translations in the Kerr spacetime now have a long history; see \cite{JanisNewman, NewmanComplex, VisserNJ}.

\subsection{Gravitational waves}

For our final example, we consider a non-stationary system, namely a gravitational plane wave. Nearly every textbook on general relativity derives the first-order metric perturbations associated with gravitational waves propagating on a flat background \cite{PoissonWill, Wald}. In transverse-traceless gauge and using background Minkowski coordinates $(t,x^i) = (t,x,y,z)$, a vacuum wave traveling in the $+z$ direction can be associated with
\begin{align}
  h_{\mu\nu}dx^\mu dx^\nu = A_+ (u) (dx^2 - dy^2) + 2 A_\times (u) dx dy,
  \label{hTT}
\end{align}
where $u \equiv t-z$ is a null phase coordinate and $A_+(u)$ and $A_\times(u)$ are arbitrary waveforms associated with the $+$ and $\times$ polarization states. As in our other examples, the metrics $\eta_{ab} + \epsilon h_{ab}$ satisfy the vacuum, $\Lambda = 0$ Einstein equation only through $O(\epsilon)$. They are not exact solutions.

It is straightforward in this case to correctly guess that the Kerr-Schild 1-form can be taken to be
\begin{equation}
  \ell_a = -\bar{\nabla}_a u,
\end{equation}
the orbits of which may be interpreted as the rays of the gravitational wave. This is clearly geodesic and tangent to the $t = z$ null hyperplanes. It is also constant. Given this, all that remains to deriving exact gravitational wave solutions is to solve \eqref{hTrans} for an appropriate gauge vector $\xi^a$ and a Kerr-Schild potential $V$.

This is considerably simpler than in the Kerr problem discussed above, and a solution can be obtained almost by inspection. The result is that
\begin{align}
	\xi^\mu \partial_\mu = - \frac{1}{4} \big[(x^2-y^2) \partial_u A_+ + 2 xy \partial_u A_\times \big] \ell^\mu dx_\mu
	\nonumber
	\\
	~ - \frac{1}{2} \big[ (x \partial_x - y \partial_y)  A_+ + (x \partial_y + y \partial_x) \big] A_\times 
\end{align}
is one possible gauge vector, and that the associated Kerr-Schild potential is
\begin{equation}
  V = \frac{1}{2} [(x^2-y^2) \partial_u^2 A_+ + 2 xy \partial_u^2 A_\times].
\end{equation}
These relations generate the metrics
\begin{equation}
  g_{ab}^{(\epsilon)} = \eta_{ab} +  \frac{\epsilon}{2} [(x^2-y^2) \partial_u^2 A_+ + 2 xy \partial_u^2 A_\times] \bar{\nabla}_a u \bar{\nabla}_a u,
  \label{gPW}
\end{equation}
which are exact solutions to the vacuum Einstein equation for all $\epsilon$ and for all seed waveforms $A_{+,\times}(u)$. These metrics were originally obtained in a completely different context \cite{Brinkmann}, and required several decades before being correctly interpreted as globally well-behaved gravitational plane waves \cite{Bondi}. Our derivation, starting from the textbook transverse-traceless approximation, makes its interpretation especially clear. The reverse procedure, deriving \eqref{hTT} as an approximation to \eqref{gPW}, is discussed in detail in \cite{HarteOptics}.

Besides being exact, it is also noteworthy that the Kerr-Schild gauge provides a potential $V$ which is more directly physical than the transverse-traceless one. The exact Riemann tensor is directly proportional to $V$, but not to $A_{+,\times}$. At linear order, coordinate transformations which amount to different choices of coordinate-fixed geodesics can be used to freely alter both of the transverse-traceless waveforms by arbitrary, linearly-growing terms. Generalizing the transverse-traceless gauge beyond linear order introduces even more complicated ambiguities \cite{HarteOptics}. The Kerr-Schild potential is instead simple to all orders and has a direct physical interpretation.

\section{Discussion}
\label{Sect:Discussion}

We have shown that if an exact solution to the linearized Einstein equation is available---obtained in any gauge---and if there exists a linearized gauge transformation which converts this to Kerr-Schild form, the resulting metric will be an exact solution to the fully-nonlinear Einstein equation. This result was used in Sect. \ref{Sect:Examples} to derive the exact Kerr and plane wave metrics from their linearized, Lorenz-gauge approximations.

Our result suggests that gauge choice can play a significant role in perturbation theory not only as something to be exploited for calculational or interpretational convenience, but also as something which can strongly affect the accuracy of a given approximation. In effect, we have advocated for the ``Kerr-Schild gauge,'' in which metric perturbations can be written in the form $V \ell_a \ell_b$ for some null $\ell_a$. This can be characterized in various ways, for example by the existence of a null 1-form $\ell_a$ such that $h_{a[b}^{\mathrm{KS}} \ell_{c]} = 0$. The Kerr-Schild gauge also satisfies \eqref{KSprops}, and can be viewed as a special kind of radiation gauge in the sense that
\begin{equation}
  h_{ab}^{\mathrm{KS}} \ell^b = \bar{g}^{ab} h_{ab}^{\mathrm{KS}} = 0.
\end{equation}
It is not however Lorenz: $\bar{\nabla}^a h_{ab}^{\mathrm{KS}} \neq 0$ in general.

Achieving the Kerr-Schild gauge is unfortunately impossible in generic systems, so the interesting question for future work is if the simplifications derived here can be generalized in some systematic way. It is unlikely that exact solutions can be easily generated in much more general settings, but perhaps errors can be significantly reduced at least for some generic class of physically-relevant systems. One possible approach is to allow for more degrees of freedom in the final gauge while also preserving the Kerr-Schild structure when it exists. This occurs naturally in the metric decomposition
\begin{equation}
  g_{ab} = \omega^2 (\bar{g}_{ab} + 2 V \ell_{(a} k_{b)}),
\end{equation}
where $\ell_a$ and $k_a$ are both null. It was shown in \cite{LlosaCarot} that generic metrics can be placed into this form with $\bar{g}_{ab}$ flat, and it was argued in \cite{HarteNonlinear} that doing so eliminates much of the nonlinearity from Einstein's equation. This also reduces to an ordinary Kerr-Schild decomposition if $\omega \rightarrow 1$ and $k_a \rightarrow \ell_a$. We briefly discussed a simple version of this generalization in Sect. \ref{Sect:Sph}, and showed that simplifications still occur at least in spherical symmetry. How far these can be extended is yet is to be determined.

\end{document}